\documentclass[aps,prb,twocolumn,showpacs,amsfonts]{revtex4}

\usepackage{amsmath}
\usepackage{graphicx}
\usepackage{multimedia}
\usepackage{color}
\usepackage{bbold}
\usepackage{bm}
\allowdisplaybreaks

\newcommand{\mb}{\mathbf}
\newcommand{\up}{\uparrow}
\newcommand{\down}{\downarrow}

\begin{document}

\title{Josephson effect between triplet superconductors through a finite ferromagnetic barrier}

\author{B. Bujnowski}
\email{bogusz.bujnowski@mailbox.tu-dresden.de}
\author{C. Timm}
\email{carsten.timm@tu-dresden.de}
\author{P. M. R. Brydon}
\email{brydon@theory.phy.tu-dresden.de}
\affiliation{Institut f\"{u}r Theoretische Physik, Technische Universit\"{a}t
Dresden, 01062 Dresden, Germany}

\date{\today}

\begin{abstract}
Charge and spin transport in a junction involving two triplet superconductors
and a ferromagnetic barrier are studied. We use
Bogoliubov-de Gennes wavefunctions to construct the Green's function, from which
we obtain the Josephson currents in terms of the Andreev
reflection coefficients. We focus on the consequences of a finite barrier width
for the occurrence of 0-$\pi$ transitions and for the spin currents, and examine
the appropriateness of the common $\delta$-function approximation for the
tunneling region.
\end{abstract}


\maketitle

\section{Introduction}

Josephson junctions between spin-singlet superconductors with ferromagnetic
tunneling barriers have recently 
attracted much interest.\cite{RevModPhys.77.1321,RevModPhys.76.411,RevModPhys.77.935} Since spin-singlet
superconductivity is strongly suppressed by the exchange splitting of
the Fermi surface in the ferromagnet, such junctions represent a unique
way of probing the interplay of two competing states. A
startling consequence is the so-called $0$-$\pi$ transition: By
varying the width of the ferromagnetic tunneling barrier or changing the
magnitude of the exchange splitting, the current vs. phase relationship
reverses sign relative to the non-magnetic
case.\cite{JETP.74.178,Tanaka1997357,PhysRevLett.98.107002,PhysRevLett.90.137003,PhysRevLett.102.227005,PhysRevB.75.094514,PhysRevLett.86.2427,PhysRevB.43.10124,PhysRevB.68.014501,PhysRevB.62.11812,PhysRevB.83.144520,NatPhys.4.138,PhysRevB.77.214506,Nature.439.825,PhysRevLett.97.247001,PhysRevLett.92.057005} 
Roughly speaking, the pair-breaking effect of the exchange fields causes a spatial
oscillation of the superconducting pairing correlations within the
ferromagnetic tunneling barrier, which can therefore connect the two
superconductors with either a $0$ or $\pi$ phase shift. This phenomenon has been
the subject of intense
theoretical\cite{RevModPhys.77.1321,RevModPhys.76.411,RevModPhys.77.935,JETP.74.178,Tanaka1997357,PhysRevLett.98.107002,PhysRevLett.90.137003,PhysRevLett.102.227005,PhysRevB.75.094514,PhysRevB.43.10124,PhysRevB.68.014501,PhysRevB.62.11812,NatPhys.4.138,PhysRevB.83.144520}  
and experimental\cite{PhysRevLett.86.2427,PhysRevB.77.214506,Nature.439.825,PhysRevLett.97.247001,PhysRevLett.92.057005} interest and is by now very well 
understood.

At least from a theoretical point of view, it is therefore surprising
that the physics of triplet superconductor (TSC)--ferromagnet (FM)
bilayers~\cite{PhysRevB.67.174501,PhysRevB.75.134510,PhysRevB.70.214524,PhysRevB.80.224520,PhysRevB.83.060508,PhysRevB.83.180504}
and junctions has received relatively little
attention.\cite{PhysRevB.75.134508,PhysRevB.75.094514,PhysRevLett.96.047009,PhysRevB.70.214524,PhysRevB.77.104504,JPSP,PhysRevLett.103.147001,springerlink:10.1007/s10909-007-9446-2,Rahnavard}
Due to the intimate relationship between triplet pairing and ferromagnetism, one
would anticipate very different behavior compared to the singlet
case.
Recently, several papers have pointed out that the orientation
of the magnetization relative to the TSC vector order
parameter is a key parameter of such systems,
leading to a variety of exotic effects: A universal $0$-$\pi$ transition
dependent upon 
the orientation of the magnetic moment, an intimate connection between
spin and charge Josephson effects, and also phase-independent 
Josephson spin
currents.\cite{PhysRevB.77.104504,JPSP,PhysRevLett.103.147001,PhysRevB.80.224520,Rahnavard,PhysRevB.83.180504} 
Although none of these predictions have yet been
verified experimentally, the recent creation of superconducting thin
films of Sr$_2$RuO$_4$ is a significant step towards the fabrication of
TSC-FM heterostructures.\cite{AppliedPhysics}
A deeper understanding of the physics of these systems is therefore
desirable.

Most investigations of TSC junctions were performed assuming an atomically thin
barrier modeled by a $\delta$-function separating the superconducting
regions. The effect of a finite barrier 
width have only been studied by Rahnavard \emph{et al.}\cite{Rahnavard} for
the case of a $(p_z+ip_y)$-wave gap where the authors obtained results that are
 only partially consistent with the previous perturbative
calculations~\cite{PhysRevLett.103.147001} for a $\delta$-function 
barrier. This raises the interesting possibility that additional
mechanisms for the Josephson currents appear due to the finite barrier
width. It is therefore the aim of this work to systematically investigate the
relationship between the results in the $\delta$-function limit and for a more
realistic barrier model. In our calculations we utilize a 
non-perturbative method to determine the Josephson current by solving the
Bogoliubov-de Gennes (BdG) equation. We compare the results with the
  predictions of perturbation theory.~\cite{PhysRevLett.103.147001}

\section{Theory}

\begin{figure}
   \begin{center} 
   \includegraphics[width=\columnwidth]{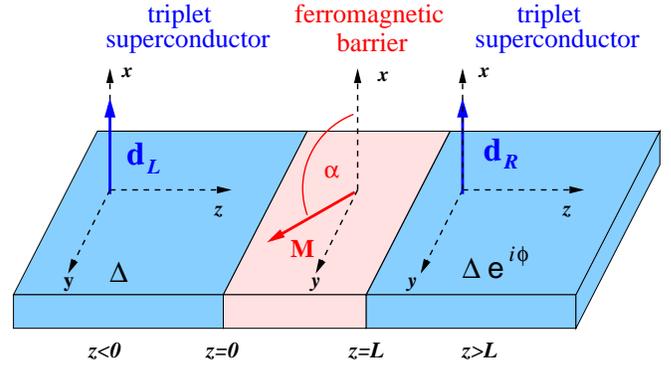}
   \caption{Schematic diagram of the junction.}\label{junction}
   \end{center}
\end{figure}

We consider a Josephson junction consisting of an FM layer of width
$L$ sandwiched between two TSCs, shown schematically
in Fig.\ \ref{junction}. The BdG equation for quasiparticle states with
energy $E$ reads
\begin{align}\label{BDG}
\left(
\begin{array}{cc}
\hat{\mathcal{H}}_0(\mb{r}) & \hat{\Delta}(\mb{r})\\
\hat{\Delta}^\dagger(\mb{r}) & -\hat{\mathcal{H}}^T_0(\mb{r})
\end{array}\right)\Psi(r)=E\Psi(r).
\end{align}
The noninteracting Hamiltonian is
\begin{align}
\hat{\mathcal{H}}_0(\mb{r})=\left[-\frac{\hbar^2}{2m}\nabla_\mb{r}^2-\mu\right]
\mathbb{1}-\Theta(z)\Theta(L-z)g\mu_B\hat{\bm{\sigma}}\cdot\mb{M} ,
\end{align}
where for simplicity we assume two-dimensional circular Fermi surfaces lying
in the $y$-$z$ plane, as well as equal effective masses and chemical potentials
in all three regions of the junction. The magnetic moment $\mb{M}$ of the FM
lies in the $x$-$y$ plane and forms an angle $\alpha$ with the $x$ axis so that
\begin{align}
\mb{M}=M\, (\cos\alpha,\sin\alpha,0).
\end{align}
This formulation is completely general since the TSCs are invariant under
spin rotations 
around the $x$ axis.

The gap matrix in equation (\ref{BDG}) is
\begin{align}
\hat{\Delta}(\mb{r})=\hat{\Delta}_L\Theta(-z)+\hat{\Delta}_R\Theta(z-L),
\end{align}
where $\hat{\Delta}_\nu$ ($\nu=L,R$) is related to the $\mb{d}$ vector by
$\hat{\Delta}_\nu=i\hat{\sigma}_y\,(\hat{\bm{\sigma}}\cdot\mb{d}_\nu)$.
We assume the TSCs to be in an equal-spin-pairing unitary state.
In particular, we take
\begin{eqnarray}
\mb{d}_L(\mb{k}) & = & \Delta(\mb{k})\mb{e}_x, \\
\mb{d}_R(\mb{k}) & = & \Delta(\mb{k})e^{i\phi}\mb{e}_x,
\end{eqnarray}
where $\Delta(\mb{k})$ denotes the orbital pairing state, so that the two
superconductors differ only in the overall phase $\phi$ of the order parameter.
In this work we will be concerned with the three orbital pairing states
\begin{align}
\Delta(\mb{k})=
\left\{\begin{array}{cl}
\Delta(T)k_y/k_F,& p_y \text{-wave}, \\
\Delta(T)k_z/k_F,& p_z \text{-wave}, \\
\Delta(T)(k_z+ik_y)/k_F,& p_z+ip_y \text{-wave}, \end{array}\right.
\end{align}
where $\Delta(T)\cong\Delta_0\sqrt{(T_C-T)/T_C}$ is the weak-coupling
temperature-dependent gap magnitude and $k_F$ is the Fermi momentum of the
normal state. 

The BdG equation is solved separately in the different regions under the
assumption of low-energy excitations, i.e., $\mu\gg E,$ $|\Delta(\mb{k})|$,
so that the electron and hole wavevectors can be regarded as approximately 
equal in magnitude, $k_e \approx k_h \approx k_F$, the so-called Andreev
approximation. The dephasing of the  
electron and hole wavefunctions in the finite FM layer constrains the validity
of this approximation to $(k_e - k_h)L \approx 2EL/\hbar v_{F} \ll 1$, where
$v_F\approx 10^6\, \mathrm{ms}^{-1}$ is the Fermi velocity
and $E \leq \max\{|\Delta_{\bf k}|\} \approx 0.1\,\mathrm{meV}$ (we assume
$T_C\approx 1\,\mathrm{K}$). This
implies that we restrict ourselves to thin layers with $Lk_F\lesssim100$.
In the TSC
we obtain plane-wave solutions 
$\psi_{\mb{k},e(h),\sigma}=\phi_{\mb{k},e(h),\sigma}e^{i\mb{k}\cdot{\bf r}}$ for
electron-like (hole-like) quasiparticles with wavevector $\mb{k}$ and spin
$\sigma$. The spinors are 
\begin{align}
\phi_{\mb{k},e,\up}=(s_{\up\mb{k}} u_{\mb{k}},0,v_{\mb{k}},0)^T,\\
\phi_{\mb{k},h,\up}=(s_{\up\mb{k}} v_{\mb{k}},0,u_{\mb{k}},0)^T,\\
\phi_{\mb{k},e,\down}=(0,s_{\down\mb{k}} u_{\mb{k}},0,v_{\mb{k}})^T,\\
\phi_{\mb{k},h,\down}=(0,s_{\down\mb{k}} v_{\mb{k}},0,u_{\mb{k}})^T,
\end{align}
where $u_{\mb{k}}=\sqrt{(E+\Omega_\mb{k})/2E}$,
$v_{\mb{k}}=\sqrt{(E-\Omega_\mb{k})/2E}$,
$\Omega_\mb{k}=\sqrt{E^2-|\Delta(\mb{k})|^2}$,
$s_{\sigma\mb{k}}=-\sigma\Delta(\mb{k})/|\Delta(\mb{k})|$, and
$|\mb{k}|=\sqrt{2m\mu/\hbar^2}=k_F$. In the FM region we obtain plane-wave
solutions $\psi^{FM}_{\mb{k}^s,e(h),s}=\phi^{FM}_{e(h),s}e^{i\mb{k}^s\cdot\mb{r}}$
for electron-like (hole-like) quasiparticles with wavevector $\mb{k}^s$. The
spinors of spin $s=\pm$ electrons and holes are 
\begin{align}
\phi^{FM}_{e,s}&=(s e^{-i\alpha}/\sqrt{2},1/\sqrt{2},0,0)^T,\\
\phi^{FM}_{h,s}&=(0,0,s e^{i\alpha}/\sqrt{2},1/\sqrt{2})^T.
\end{align}
Due to the exchange splitting in the FM, we have majority-spin and minority-spin
Fermi surfaces with radii
\begin{align}
|\mb{k}^s|=k_F^s=\sqrt{\frac{2m}{\hbar^2}(\mu+sg\mu_BM)}=k_F\sqrt{1+s\lambda} ,
\end{align}
where $\lambda$ denotes the ratio of the exchange splitting to the chemical
potential.

\begin{figure*}
   \begin{center} 
    \includegraphics[width=4.5in]{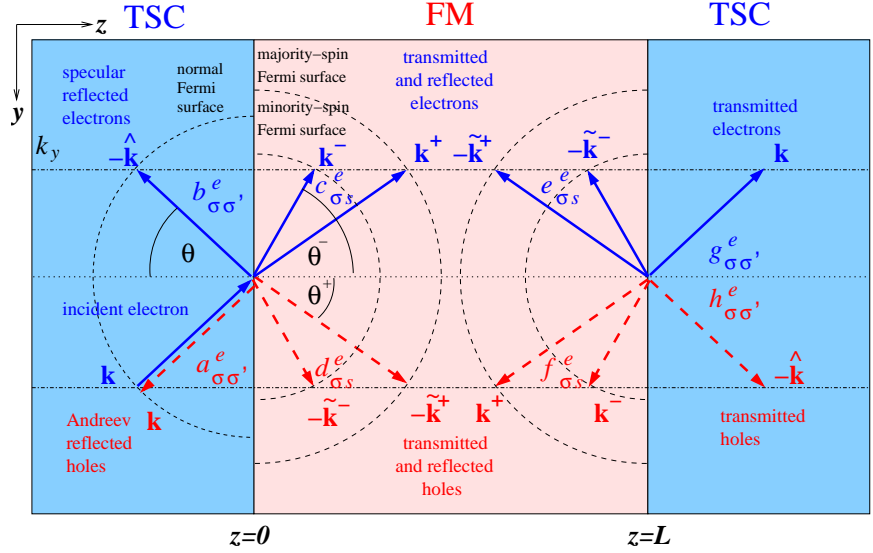}
    \caption{(Color online) Schematic representation of the various wave
      vectors, 
    amplitudes, and coefficients for a spin-$\sigma$ electron-like quasiparticle
    incident from the left TSC, with wavefunction
    $\Psi_{e,\sigma}(\mb{r})$. The arrows denote the directions of travel of
    electron-like quasiparticles (solid blue) and hole-like quasiparticles
    (dashed red).\label{scatter}}
   \end{center}
\end{figure*}

We use these piecewise solutions to construct the wavefunction ansatz for a
spin-$\sigma$ electron-like quasiparticle incident from the left TSC with
wavevector $\mb{k}=(k_y,k_z)=k_F(\sin\theta,\cos\theta)$, where $\theta$ is the
injection angle relative to the $z$ axis:
\begin{align}\label{elinj}
&\Psi_{e,\sigma}(\mb{r})=\notag\\ &\Theta(-z)\Big\{\psi^L_{\mb{k},e,\sigma}+\sum_{\sigma^\prime=\up\down}\big[a^e_{\sigma\sigma^\prime}\psi_{\mb{k},h,\sigma^\prime}^L+b^e_{\sigma\sigma^\prime}\psi^L_{-\hat{\mb{k}},e,\sigma^\prime}\big]\Big\}\notag\\
&+\Theta(z)\Theta(L-z)\Big\{\sum_{s=+-}\big[c^e_{\sigma{s}}\psi^{FM}_{\mb{k}^s,e,s}+d^e_{\sigma{s}}\psi^{FM}_{-\widetilde{\mb{k}}^s,h,s}\big.\notag\\&+\big.e^e_{\sigma{s}}\psi^{FM}_{-\widetilde{\mb{k}}^s,e,s}+f^e_{\sigma{s}}\psi^{FM}_{\mb{k}^s,h,s}\big]\Big\}\notag\\
&+\Theta(z-L)\Big\{\sum_{\sigma^\prime=\up\down}[g^e_{\sigma\sigma^\prime}\psi^R_{\mb{k},e,\sigma^\prime}+h^e_{\sigma\sigma^\prime}\psi^R_{-\hat{\mb{k}},h,\sigma^\prime}]\Big\}.
\end{align}
A schematic representation of the wavevectors and transmission and reflection amplitudes appearing in the wavefunction $\Psi_{e,\sigma}(\mb{r})$ is
shown in Fig.\ \ref{scatter}. For $z<0$ this ansatz describes Andreev
reflection of hole-like quasiparticles 
with wavevector $\mb{k}$ and specular reflection of electron-like quasiparticles
with wavevector $-\hat{\mb{k}}=(k_y,-k_z)$. These processes are weighted by
their probability amplitudes $a^e_{\sigma\sigma^\prime}$ and
$b^e_{\sigma\sigma^\prime}$, respectively. In the FM (i.e., for $0<z<L$), the
wavevectors of the transmitted and reflected quasiparticles are spin-dependent.
For wavevectors of spin-$s$ right-moving (left-moving) electrons (holes) 
we write $\mb{k}^s=(k_y,k_z^s)$, while spin-$s$ left-moving (right-moving) 
electrons 
(holes) have wavevectors $-\widetilde{\mb{k}}^s=(k_y,-k_z^s)$. The associated
probability amplitudes for electrons (holes) are $c^e_{\sigma s}$ and
$e^e_{\sigma s}$ ($d^e_{\sigma s}$ and $f^e_{\sigma s}$). For $z>L$ we have
transmitted electron-like and hole-like quasiparticles with wavevectors $\mb{k}$
and $-\hat{\mb{k}}$, and probability amplitudes $g^e_{\sigma\sigma^\prime}$ and 
$h^e_{\sigma\sigma^\prime}$, respectively. The ansatz for an
incident hole-like quasiparticle $\Psi_{h,\sigma}(\mb{r})$ is analogous.
The probability amplitudes for this case are distinguished by the superscript
$h$, 
e.g., $a^h_{\sigma\sigma^\prime}$, $b^h_{\sigma\sigma^\prime}$ etc. Due to the
translational invariance along the $y$ axis, the wavevector
component parallel to the interface is conserved during scattering, which gives
\begin{align}
k_F\sin\theta = k_F^s\sin\theta^s ,
\end{align}
where $\theta^s$ is the transmission angle for quasiparticles from the spin-$s$
Fermi surfaces. Making use of this relation, $k_z^s$ can be expressed as
\begin{align}
k_z^s=k_F^s\cos\theta^s=k_F\sqrt{\cos^2\theta+s\lambda}.
\end{align}
For injection angles
\begin{align}\label{critangle}
\theta>\theta_\text{crit}=\arccos\sqrt{\lambda} ,
\end{align}
$k_z^s$ is imaginary for minority-spin quasiparticles, i.e., the wave decays
exponentially into the FM region.

The probability amplitudes for the various processes are calculated from
the continuity of the wavefunction and its derivative at the interfaces,
\begin{align}\label{boundarystart}
\Psi_{e(h),\sigma}(\mb{r})|_{z=0^-}&=\Psi_{e(h),\sigma}(\mb{r})|_{z=0^+},\\
\Psi_{e(h),\sigma}(\mb{r})|_{z=L+0^-}&=\Psi_{e(h),\sigma}(\mb{r})|_{z=L+0^+},\\
\frac{\partial}{\partial
  z}\Psi_{e(h),\sigma}(\mb{r})\Big|_{z=0^-}&=\frac{\partial}{\partial
  z}\Psi_{e(h),\sigma}(\mb{r})\Big|_{z=0^+},\\ 
\frac{\partial}{\partial
  z}\Psi_{e(h),\sigma}(\mb{r})\Big|_{z=L+0^-}&=\frac{\partial}{\partial
  z}\Psi_{e(h),\sigma}(\mb{r})\Big|_{z=L+0^+}\label{boundaryend}. 
\end{align}
The results for a $\delta$-barrier can be obtained by taking the limits
$L\rightarrow0$, $\lambda\rightarrow \infty$ while keeping the product
$L\lambda$ constant.

The charge current $I_C$ and spin current $I_S^\mu$ ($\mu=x,y,z$) perpendicular to
the junction interface are obtained using the Furusaki-Tsukada
formulas.\cite{Furusaki,Kasiwaya_Tanaka,Asano} 
These are written in terms of the Andreev reflection coefficients
$a^e_{\sigma\sigma^\prime}$ and $a^h_{\sigma\sigma^\prime}$, where the energy
argument is analytically continued to Matsubara frequencies [$E\rightarrow
i\omega_n = i\, (2n-1)\pi/\beta$],
\begin{align}\label{current_pwave}
I_{C}=&\frac{e}{2\hbar\beta}\int dk_{y}\sum_{\omega_n}\sum_{\sigma} \notag\\
&\times\left\{
\frac{|\Delta_\mathbf{k}|}{\Omega_{n,\mathbf{k}}}
a^e_{\sigma\sigma}(\mathbf{k},\omega_n)-\frac{|\Delta_{-\hat{\mathbf{k}}}|}{\Omega_{n,-\hat{\mathbf{k}}}} a^h_{\sigma\sigma}(-\hat{\mathbf{k}},\omega_n) \right\},\\
{I_{S}^y}=&\frac{i}{4\beta}\int
dk_{y}\sum_{\omega_n}\sum_{\sigma}\sigma \notag \\& \times\left\{
\frac{|\Delta_\mathbf{k}|}{\Omega_{n,\mathbf{k}}}
a^e_{\sigma\overline{\sigma}}(\mathbf{k},\omega_n) -\frac{|\Delta_{-\hat{\mathbf{k}}}|}{\Omega_{n,-\hat{\mathbf{k}}}} a^h_{\sigma\overline{\sigma}}(-\hat{\mathbf{k}},\omega_n) \right\},\\
I_{S}^z=&-\frac{1}{4\beta}\int
dk_{y}\sum_{\omega_n}\sum_{\sigma}\sigma \notag\\
&\times\left\{
\frac{|\Delta_\mathbf{k}|}{\Omega_{n,\mathbf{k}}}
a^e_{\sigma\sigma}(\mathbf{k},\omega_n) -\frac{|\Delta_{-\hat{\mathbf{k}}}|}{\Omega_{n,-\hat{\mathbf{k}}}} a^h_{\sigma\sigma}(-\hat{\mathbf{k}},\omega_n) \right\},
\end{align}
where $\Omega_{n,\mb{k}}=\sqrt{\omega_n^2+|\Delta_\mb{k}|^2}$,
$\overline{\sigma}=-\sigma$ and $\int
dk_{y}$ denotes the integration over all momenta parallel to the
interface.~\cite{meaculpa} The $x$ component of the spin current vanishes
since the Cooper-pair spin is always perpendicular to the $\mb{d}$-vector. 

\begin{figure*}
  \begin{center}\includegraphics[clip,width=2\columnwidth]{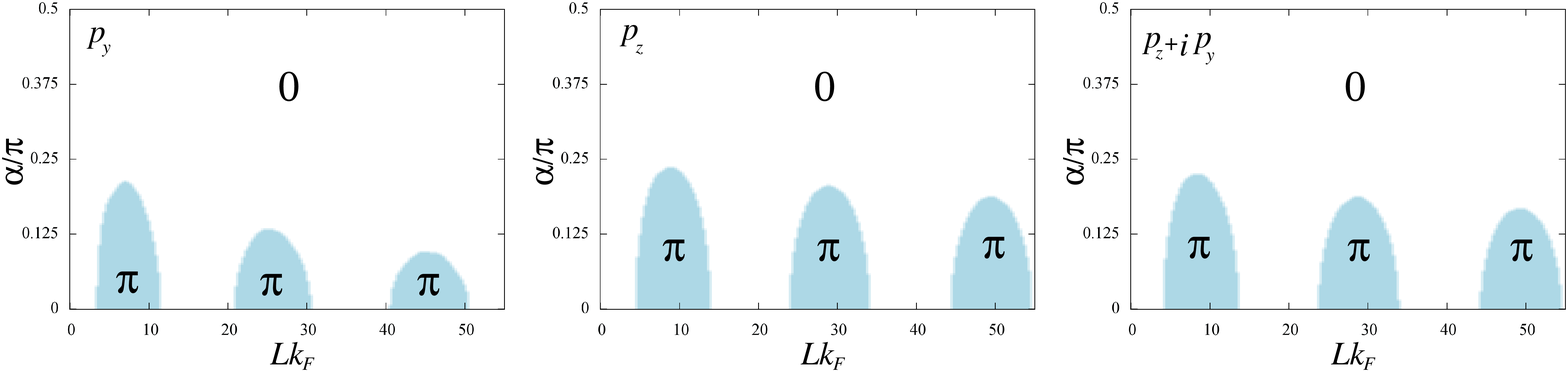}\end{center}
   \caption{Phase diagram of the junction showing $0$- and $\pi$-states as a
     function of the angle 
   $\alpha$ and barrier width $L$ for the three considered gap symmetries.
   In all panels we set $T=0.4\,T_C$ and $\lambda=0.3$.}\label{opimaps1}

   \begin{center}
   \includegraphics[clip,width=2\columnwidth]{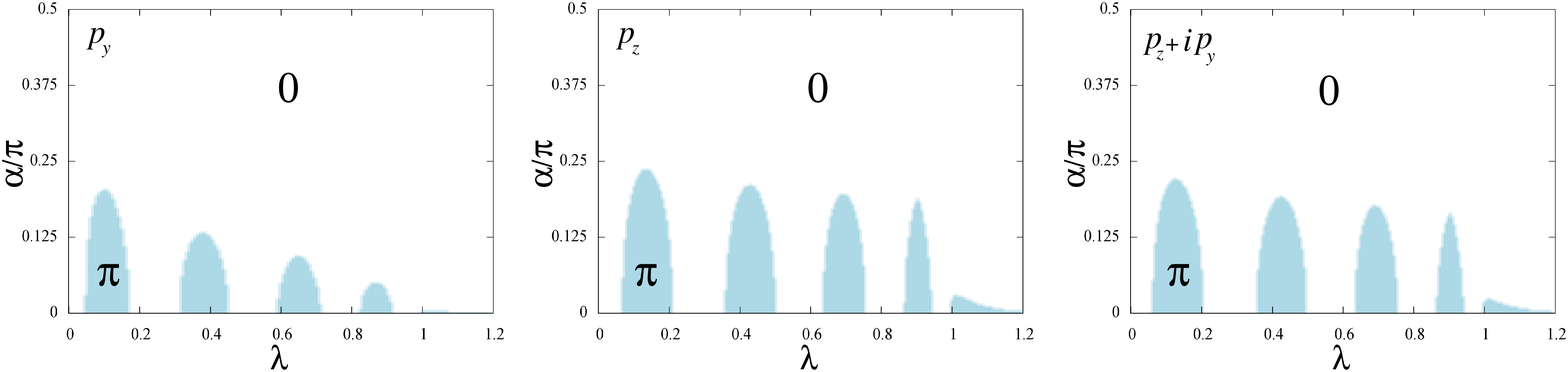} 
   \end{center}
   \caption{Phase diagram of the junction showing $0$- and $\pi$-states as a function of the angle
   $\alpha$ and exchange splitting $\lambda$ for the three considered gap
   symmetries. In all panels we set $T=0.4\,T_C$ and
   $L=20k_F^{-1}$.}\label{opimaps2}

   \begin{center}
   \includegraphics[clip,width=2\columnwidth]{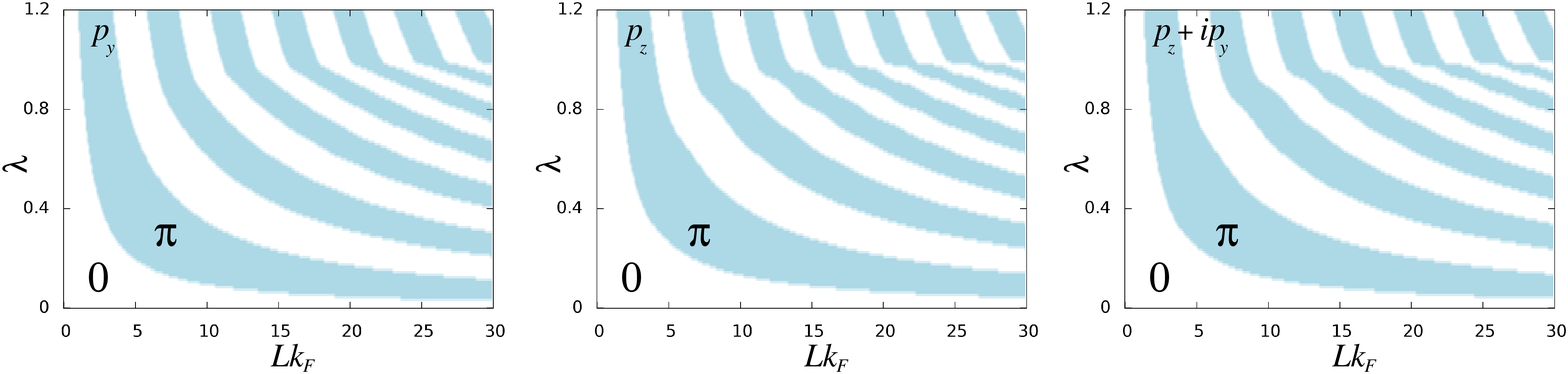}
   \end{center}
   \caption{Phase diagram of the junction showing $0$- and $\pi$-states as a function of the
   exchange splitting $\lambda$ and the barrier width $L$ for the three considered
   gap symmetries. In all panels we set $T=0.4\,T_C$ and $\alpha=0$.}\label{opimaps3}

\end{figure*}

\section{Results}

We have numerically solved Eqs.\ (\ref{boundarystart})--(\ref{boundaryend}) for
the Andreev reflection coefficients at $T=0.4\,T_C$. At least $400$ Matsubara
frequencies have been used in all calculations.

\subsection{Charge current}

\begin{figure*}[t]
   \begin{center}
    \includegraphics[clip,scale=0.3]{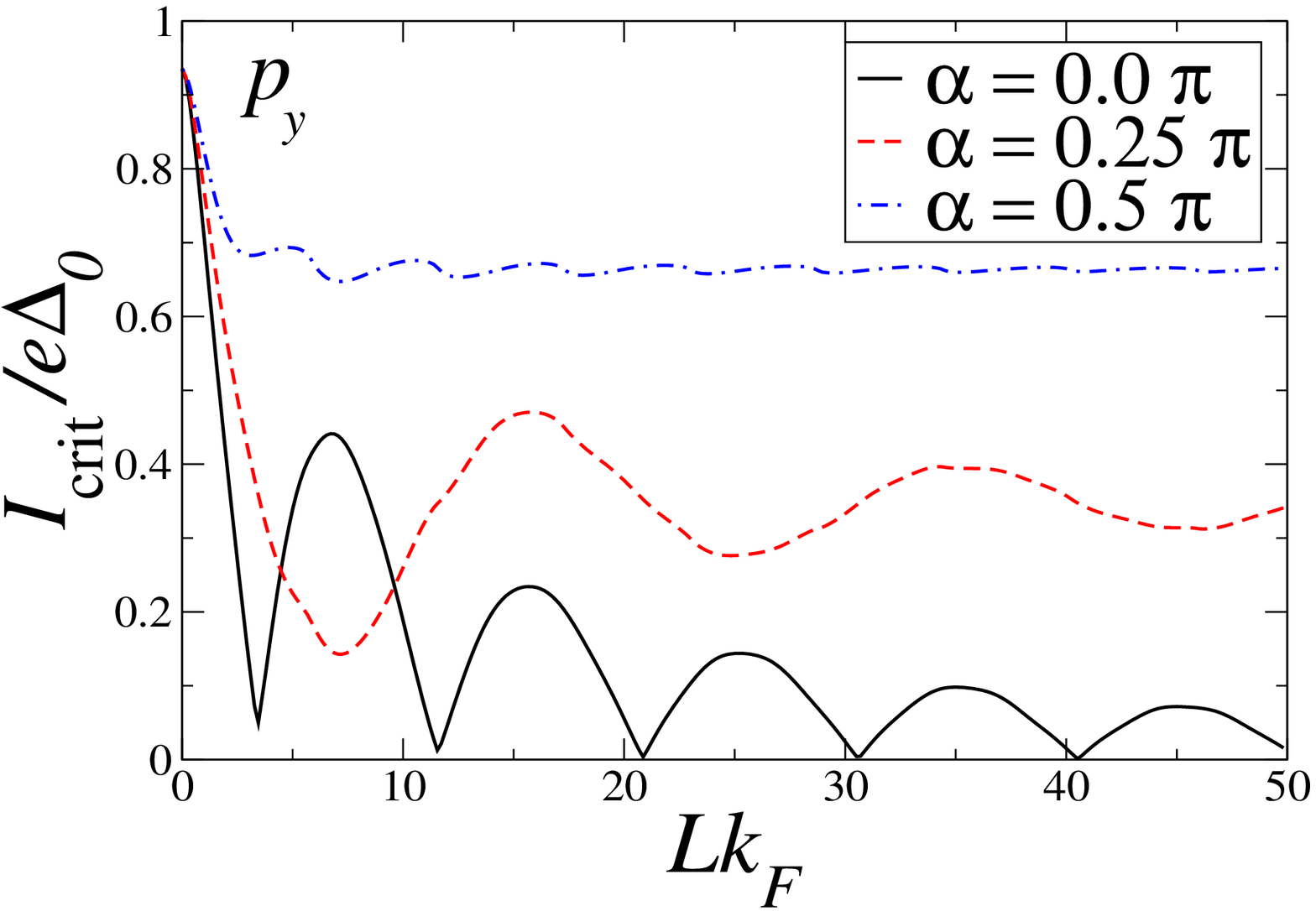}
    \includegraphics[clip,scale=0.3]{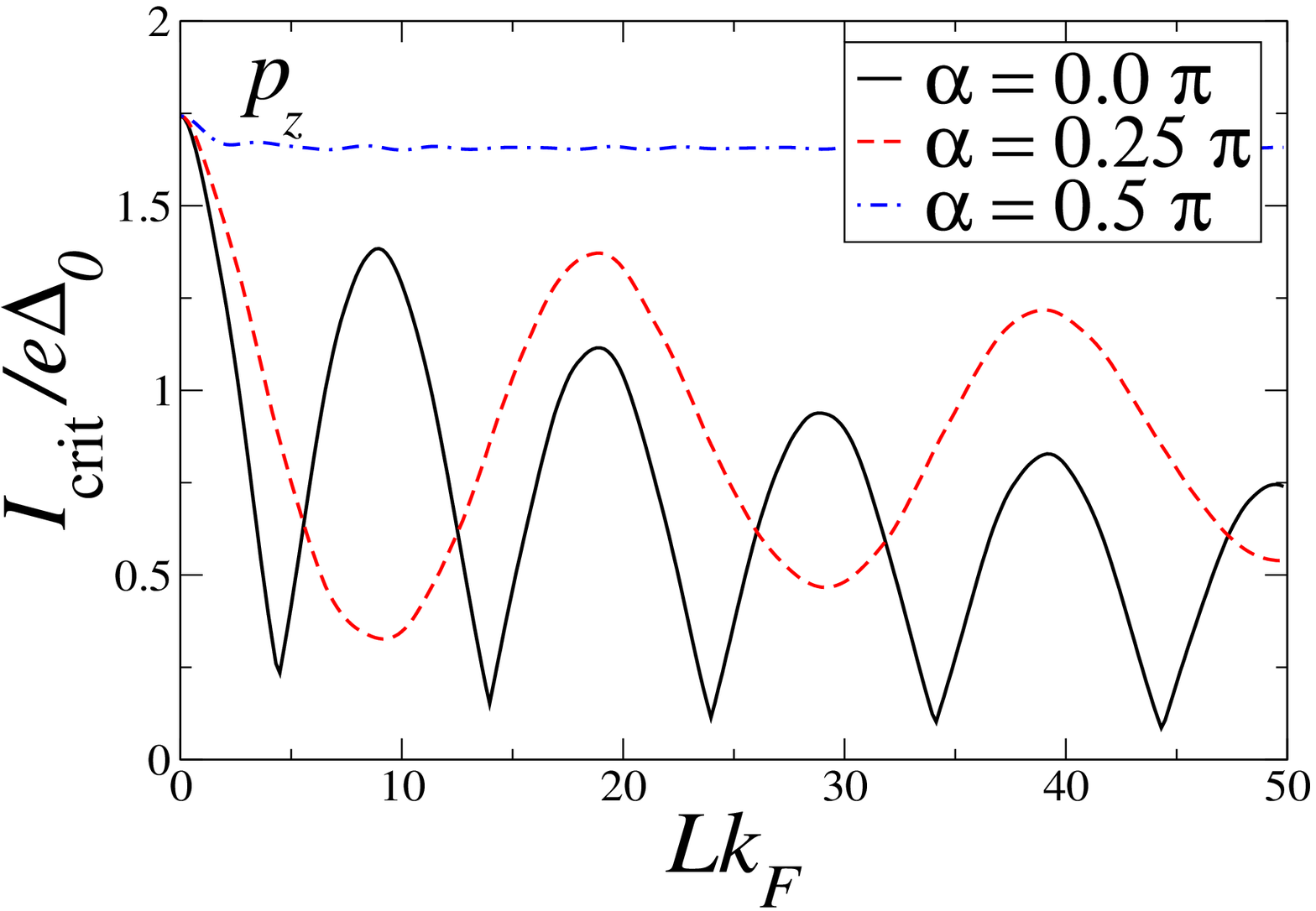}
    \includegraphics[clip,scale=0.3]{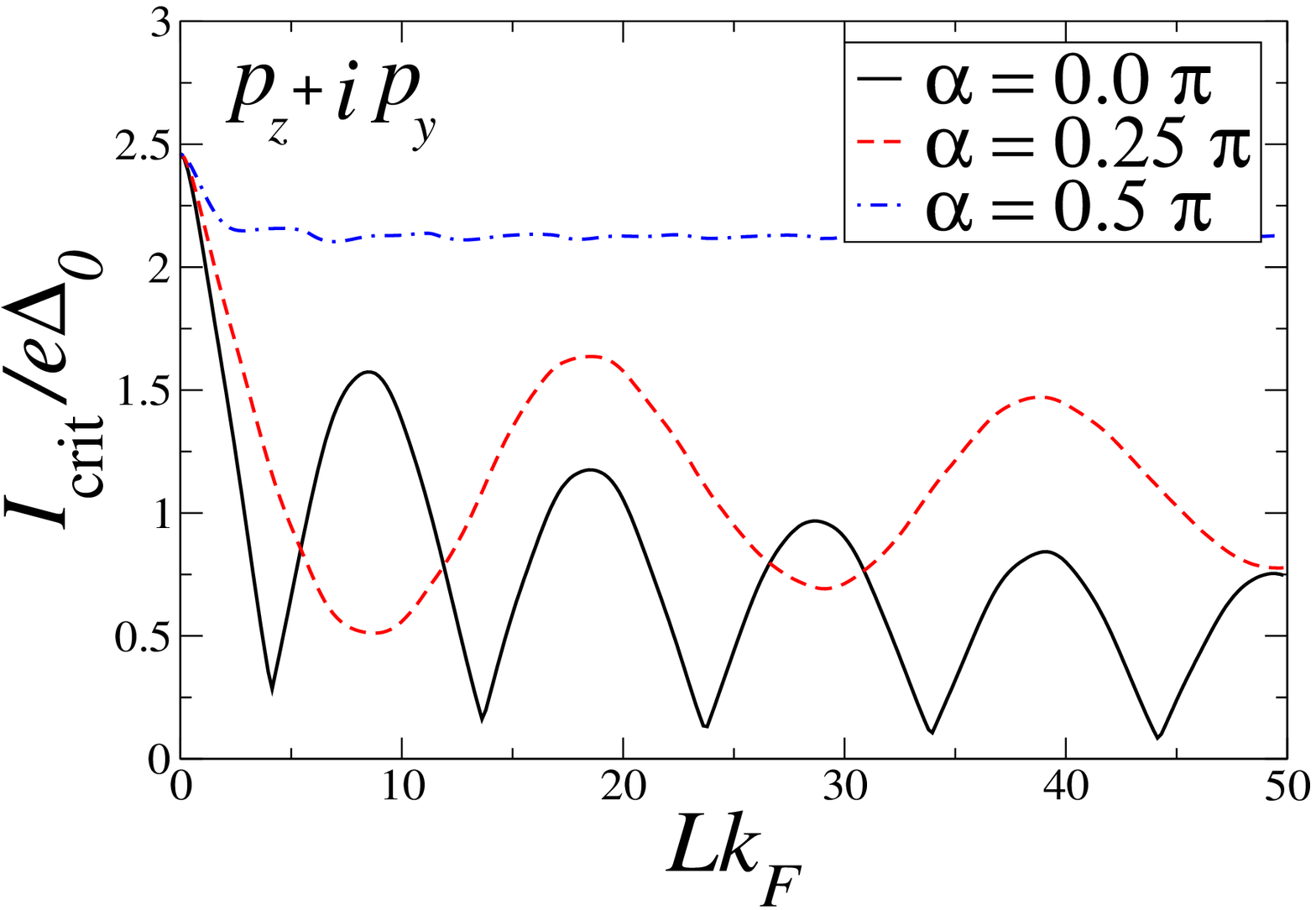}
    \end{center}
    \caption{Critical charge current as a function of the barrier width $L$. In
    all panels we set $T=0.4\,T_C$ and $\lambda=0.3$.
    }\label{critcurrent1}

    \begin{center}
    \includegraphics[scale=0.3]{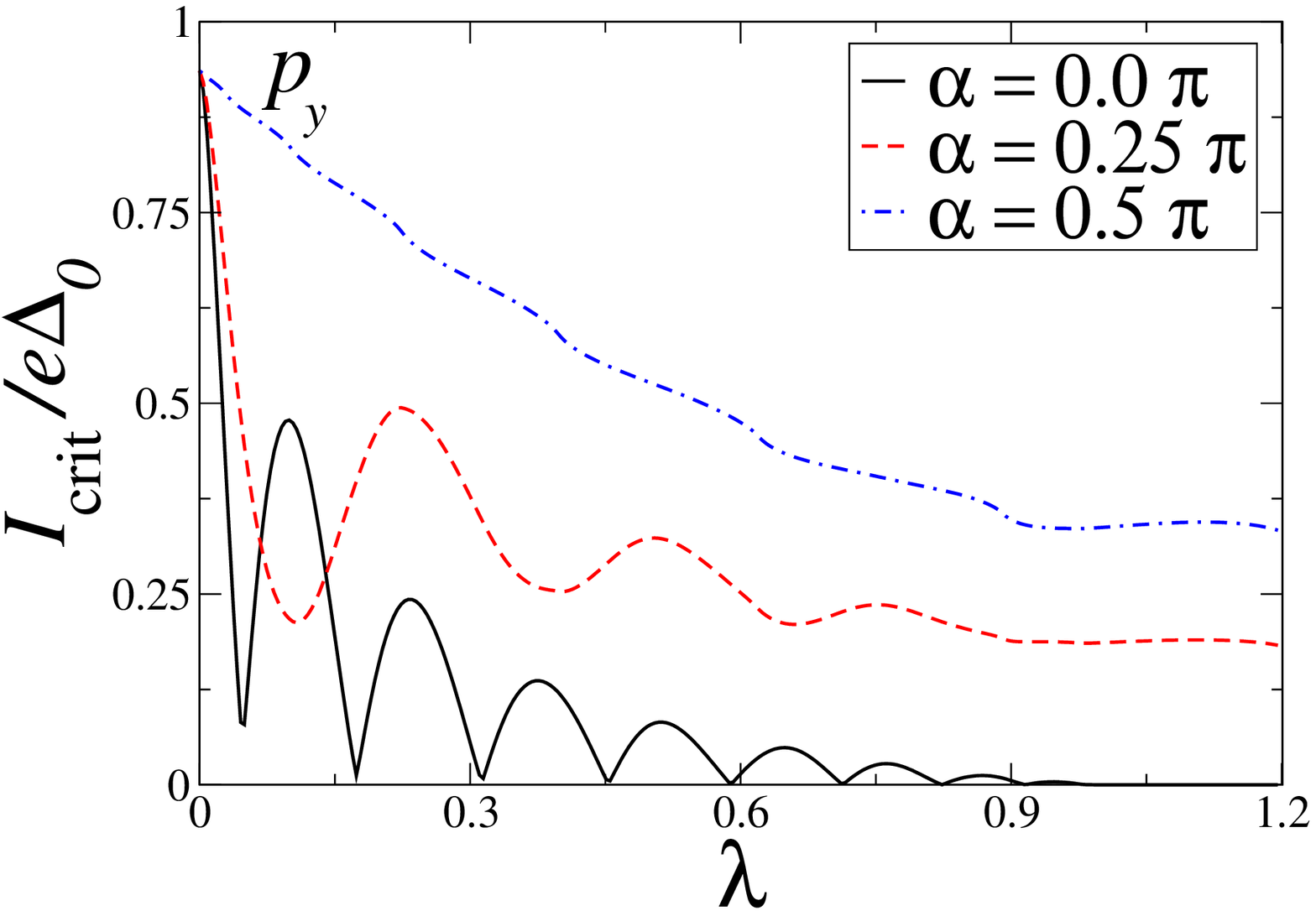}
    \includegraphics[scale=0.3]{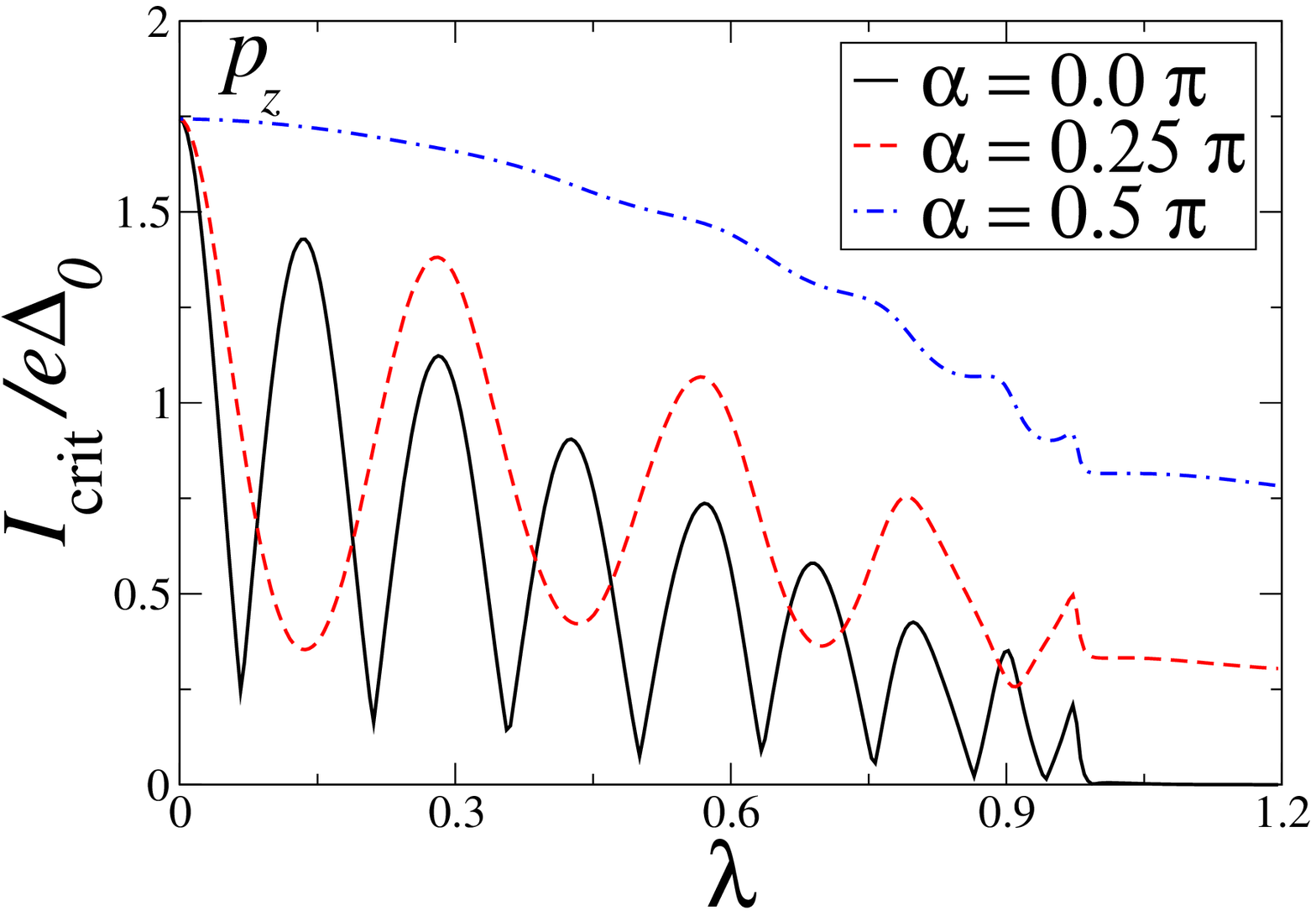}
    \includegraphics[scale=0.3]{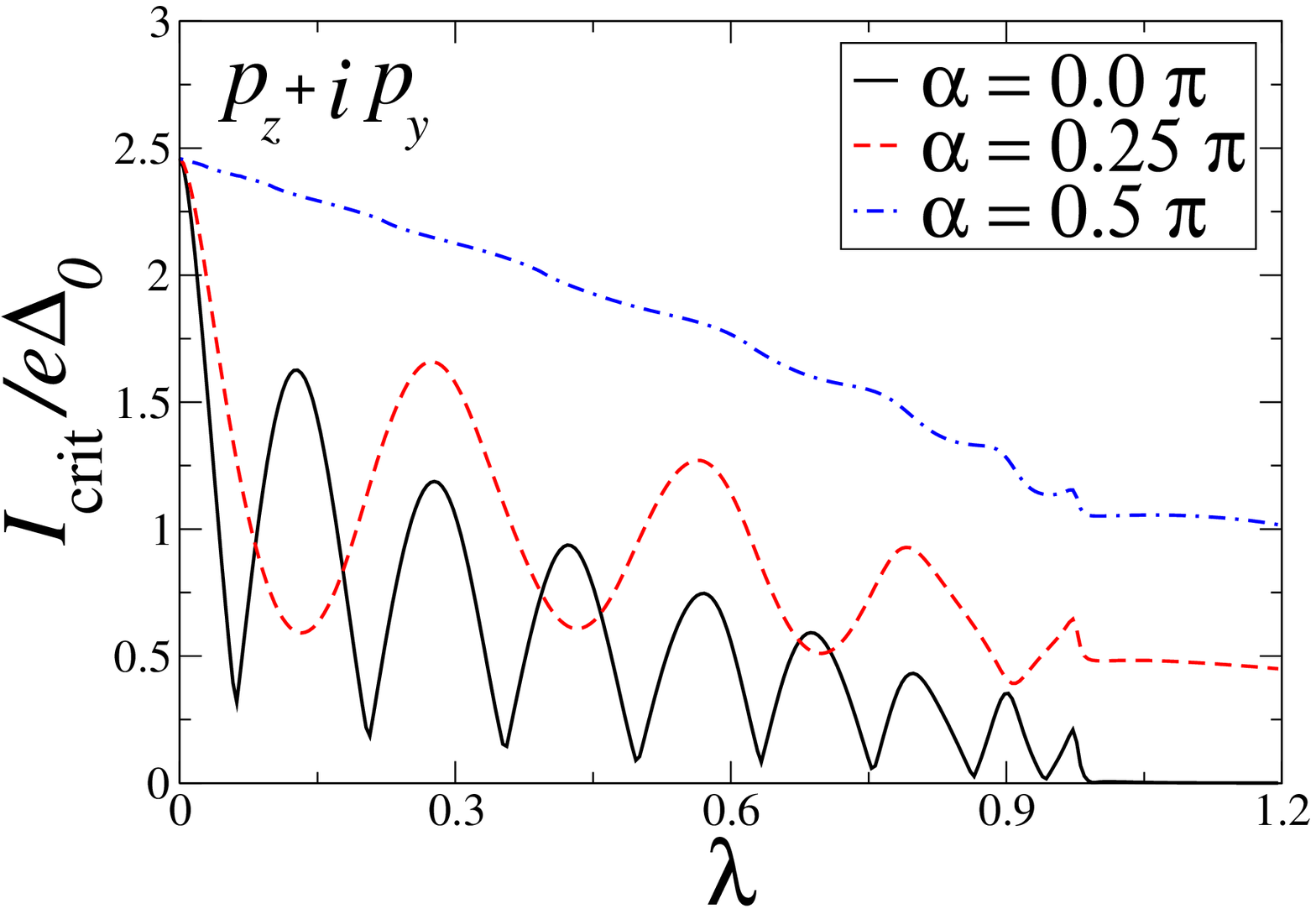}
    \end{center}
    \caption{Critical charge current as a function of the exchange splitting $\lambda$.
    In all panels we set $T=0.4\,T_C$ and
    $L=20\,k_F^{-1}$.}\label{critcurrent2}

    \begin{center}
    \includegraphics[scale=0.3]{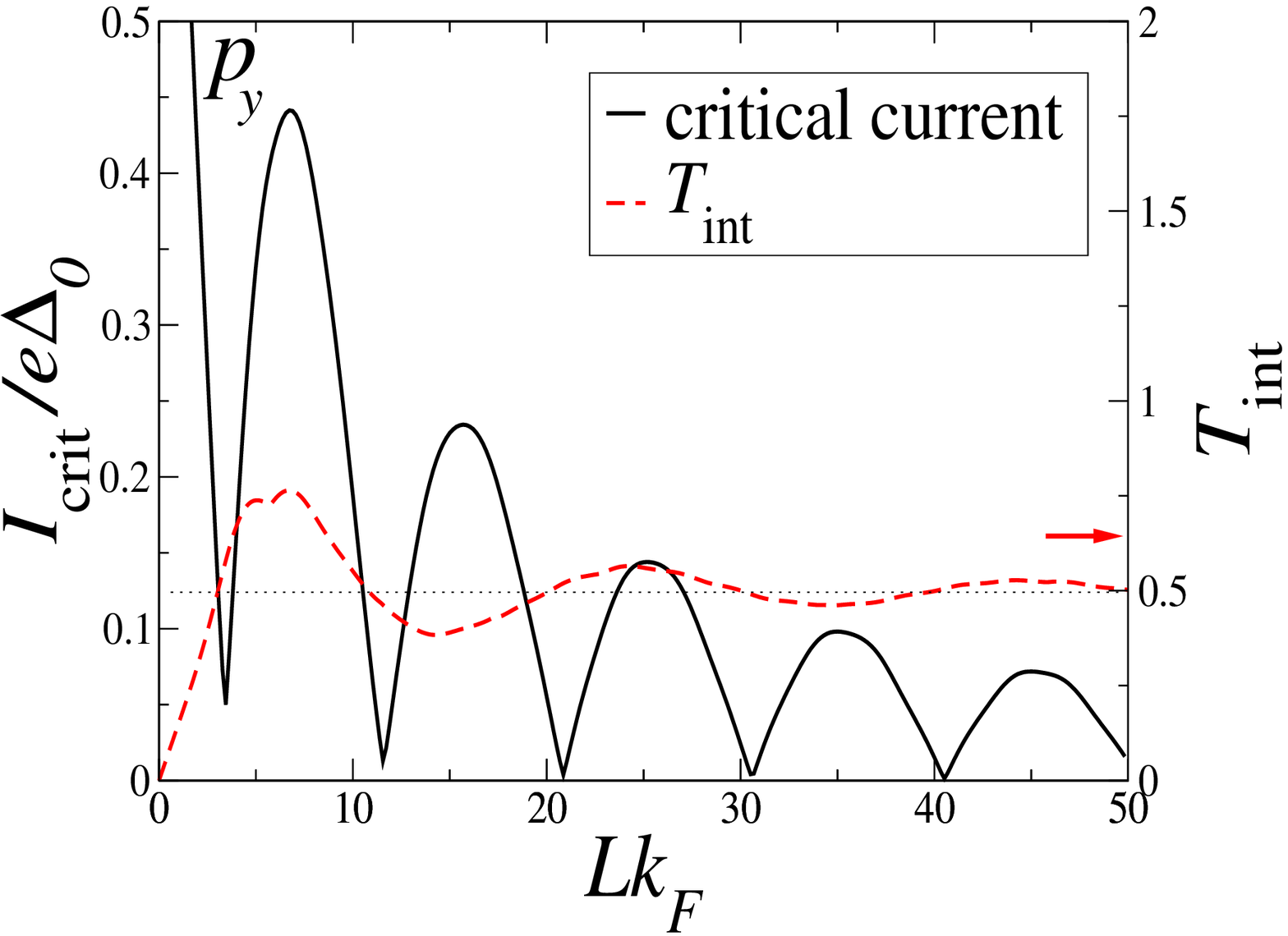}
    \includegraphics[scale=0.3]{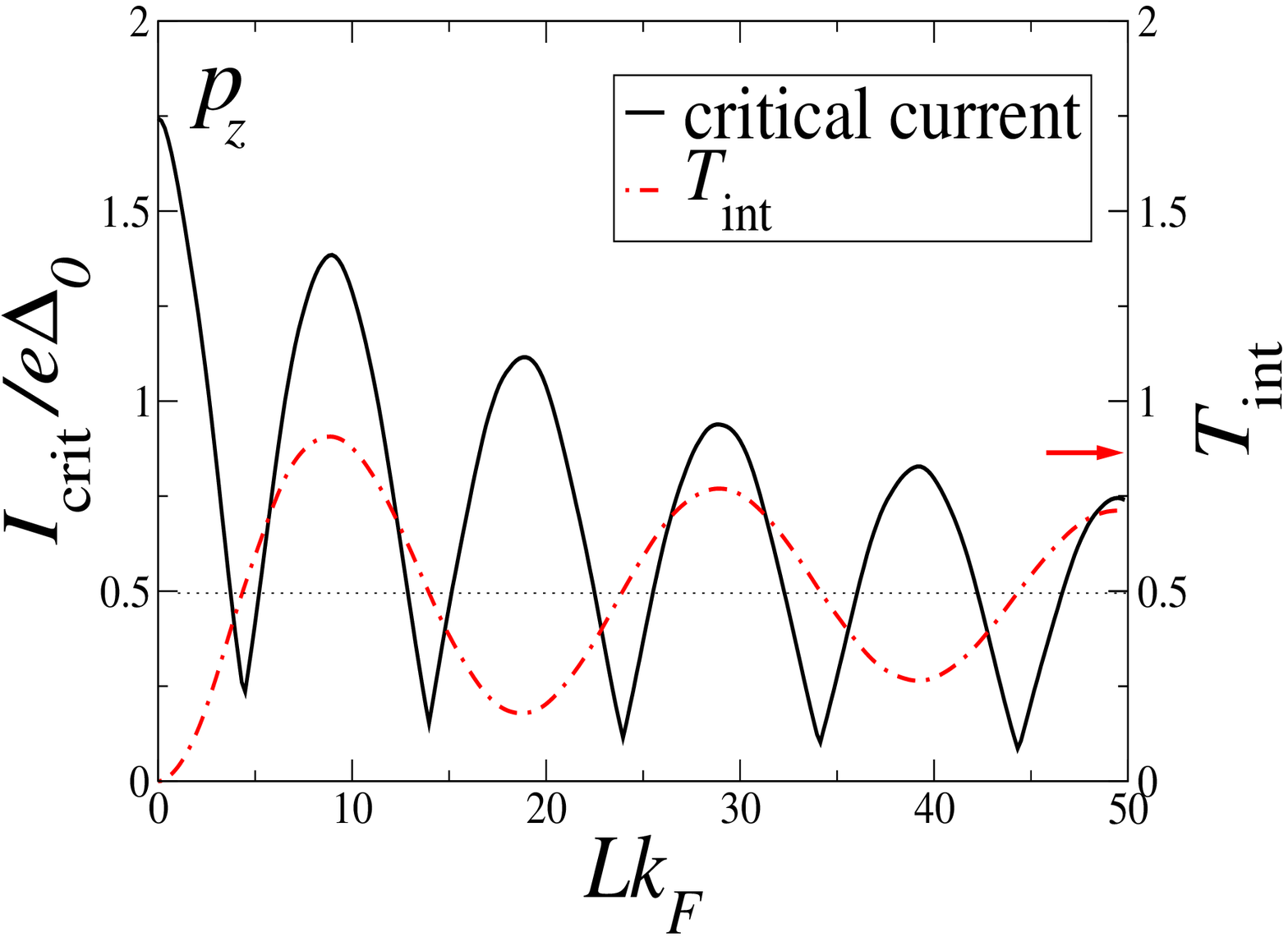}
    \includegraphics[scale=0.3]{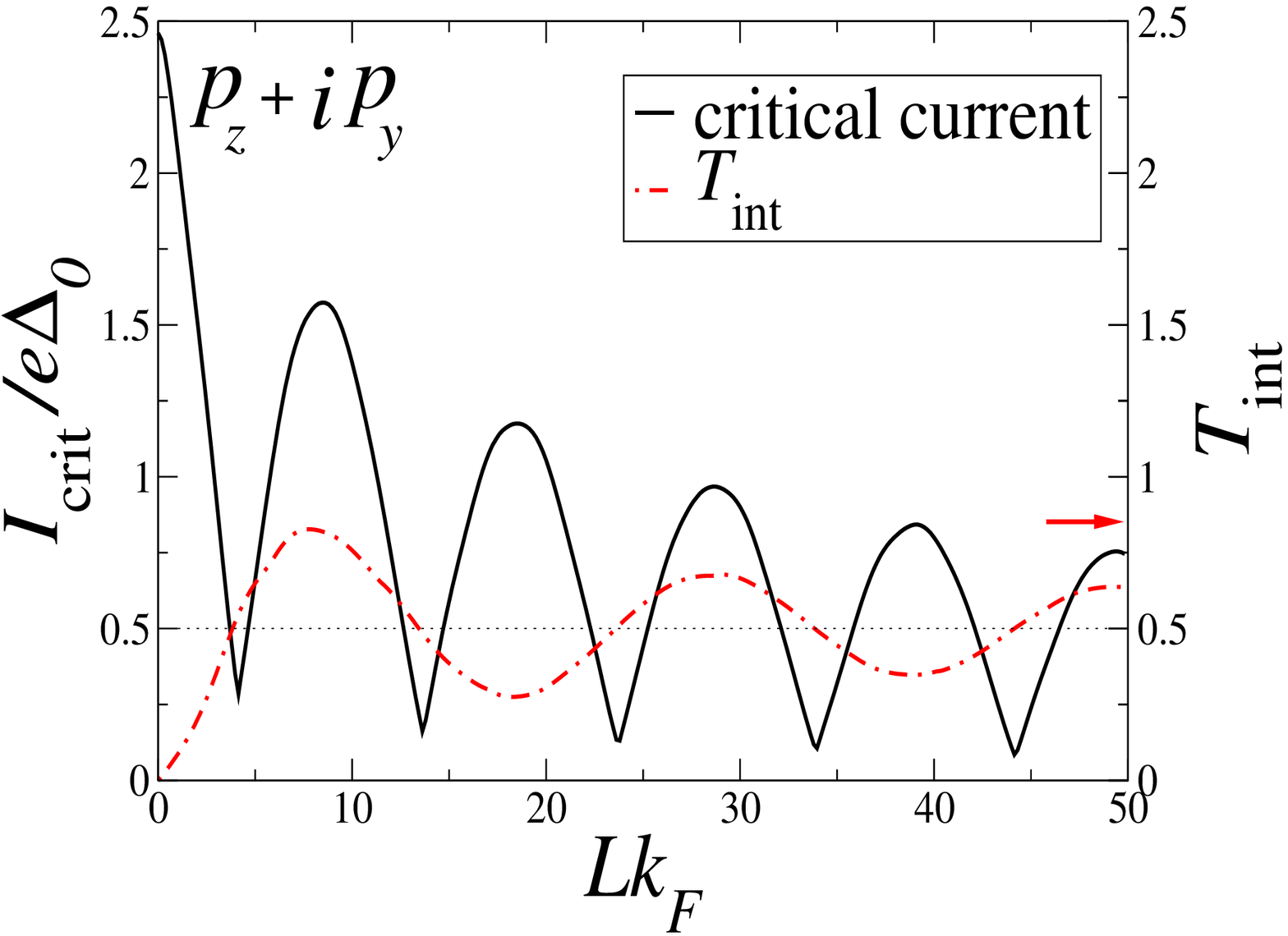}
    \end{center}
    \caption{Critical charge current (black solid lines, left axis)
    and integrated transmissivities $T_\text{int}$ (red dashed lines, right
    axis) as functions of the barrier width $L$. In all panels
    we set $\alpha=0$, $T=0.4\,T_C$, and
    $\lambda=0.3$.}\label{critcurrentweight}
\end{figure*}

We are mainly interested in the occurrence of $0$-$\pi$ transitions, which
involve a change of the phase difference $\phi$ of the global free-energy
minimum 
between $\phi=0$ and $\phi=\pi$. Points in parameter space can be identified as
corresponding to a $0$-state or a
$\pi$-state by examining the current-vs.-phase relationship. Specifically, we
have the thermodynamic relation\cite{Tinkham}
\begin{align}
I_C(\phi)=\frac{2e}{\hbar}\frac{\partial F}{\partial \phi}
\end{align}
so that $I_C=0$ implies an extremum of the free energy $F$. In our system,
minima only occur at $\phi=0$ or $\pi$. We therefore numerically calculate the
derivative of $I_C(\phi)$ at these two points to find out which one is a
minimum. If both are minima we integrate $I_C(\phi)$ to obtain the
free-energy difference $\Delta F$ between $\phi=0$ and $\pi$,
\begin{align}
\Delta F = \frac{\hbar}{2e}\int_0^\pi d \phi\, I_C(\phi) ,
\end{align}
which allows us to identify the global free-energy minimum.
Using this method, we construct phase diagrams of the junction for cuts
through the $(\alpha,\lambda,L)$ parameter space, which are presented in
Figs.~\ref{opimaps1}--\ref{opimaps3}.

We find that $0$-$\pi$ transitions are possible for all considered gap
symmetries by increasing the magnitude of the magnetization, rotating the
magnetization within the $x$-$y$ plane, or by increasing the barrier width
$L$. For 
$\lambda<1$ in the $(\alpha,\lambda)$ maps and in the $(\alpha,L)$ maps, we find
multiple dome-shaped regions where the junction is in a $\pi$-state. These
$\pi$-state domes are only found for $\alpha<\pi/4$ and their appearance is
shifted to lower values of $\alpha$ by increasing $L$ or $\lambda$. This
effect is strongest for the $p_y$-wave case, while the $\pi$ states are much
more robust for $p_z$-wave and $(p_z+ip_y)$-wave pairing.
In the $(\alpha,\lambda)$ maps the $\pi$-states are strongly suppressed
for $\lambda>1$.

For $\alpha=0$, the distribution of $0$- and $\pi$-states in
Fig.\ \ref{opimaps3} shows a  stripe-like pattern, which does not strongly
depend on the pairing symmetry of the TSC but rather upon the characteristics
of the barrier. In particular, the sharp bend of the stripes at $\lambda=1$ in
all panels is a signature of the
onset of half-metalicity since $|\mb{k}_F^-|$ vanishes at
$\lambda=1$. Interestingly, these phase diagrams are
very similar to those obtained for a junction involving
\emph{singlet} superconductors on both sides of the
barrier.\cite{PhysRevB.83.144520} The parallels between these different junctions originate from the  universal pair-breaking effect of the FM-layer. Note that for the TSC case the FM layer is
purely pair-breaking only for $\alpha=0$, and pair-breaking is completely suppressed upon
increasing $\alpha$ to $\alpha=\pi/2$. In contrast, for a
singlet-superconductor junction the FM layer is purely pair-breaking for all $\alpha$.
A $0$-$\pi$ transition caused by varying the orientation of
the magnetic moment is therefore an unambiguous experimental signature of
triplet superconductivity.

The direct observation of $\pi$-junctions is not trivial, as it requires
the detection of half vortices at the barrier.~\cite{PhysRevLett.92.057005,PhysRevB.77.214506} A more practical experimental
signature is provided by the critical charge current
\begin{align}
I_\text{crit}=\text{max}\left\{|I_C(\phi)|\right\},\;\;\phi\in[0,\pi]\, .
\end{align}
As the junction is tuned through a $0$-$\pi$ transition, the sign change of
the dominant $\sin\phi$ term in the charge current $I_C(\phi)$ causes the
appearance of sharp, cusp-like minima where $I_\text{crit}$ nearly
vanishes.~\cite{JETP.74.178} If $I_C(\phi)$ were exactly proportional to
$\sin\phi$, this sign 
change would exactly coincide both with $I_\text{crit}=0$ and with a $0$-$\pi$
transition; the presence of weak higher harmonics in $I_C(\phi)$ complicates
the analysis, but the minima in $I_\text{crit}$ nevertheless remain a good
proxy for the $0$-$\pi$ transition.
We plot $I_\text{crit}$ as a function of $L$ and $\lambda$ for $\alpha=0$,
$\pi/4$, and $\pi/2$ in Figs.\ \ref{critcurrent1} and \ref{critcurrent2}. The
sharp minima in the critical current characteristic of the $0$-$\pi$
transition are only found at $\alpha=0$, and are in good agreement with the
phase boundaries in Figs.\ \ref{opimaps1} and \ref{opimaps2}.
Upon increasing $\alpha$, the sharp minima disappear by
$\alpha=\pi/4$ and are replaced by broad minima
in $I_{\text{crit}}$, which coincide with every second maximum of the $\alpha=0$
critical current. These minima do not correspond to sign changes of
the $\sin\phi$ term. At $\alpha=\pi/2$, most structure in $I_\text{crit}$ has
vanished. 

The charge
current found by perturbation theory for a $\delta$-barrier
junction is,\cite{PhysRevLett.103.147001}
\begin{align}\label{pertubation_current}
I_C^\text{pert}(\phi)\propto\left(T_\mathrm{sp}-T_\mathrm{sf}\cos 2\alpha
\right) \,\sin\phi ,
\end{align}
where $T_\mathrm{sp}$ and $T_\mathrm{sf}$ denote the positive tunneling matrix
elements for spin-preserving tunneling and spin-flip tunneling, respectively.
As one can see, a $\pi$-state can only be realized if
$T_\mathrm{sp}<T_\mathrm{sf}$ and $\alpha<\pi/4$, with the latter condition
clearly consistent with the phase diagrams and critical current plots.

It is interesting to see if Eq.\ (\ref{pertubation_current}) is also
valid in the finite-barrier case if we regard $T_\mathrm{sp}$
and $T_\mathrm{sf}$ as functions of $L$ and $\lambda$. The matrix
elements $T_\mathrm{sp}$ and $T_\mathrm{sf}$ are expected to be 
roughly proportional to the corresponding
transmissivities.\cite{PhysRevLett.103.147001} Note that the transmissivities
are 
independent of $\alpha$ for both the $\delta$-function and finite width
barriers. From Eq.\ (\ref{pertubation_current}) we see that the spin-flip
transmissivity $\mathcal{T}_\mathrm{sf}$ should be larger than the 
spin-preserving transmissivity $\mathcal{T}_\mathrm{sp}$ in a $\pi$-state,
whereas the opposite should be 
true in a $0$-state. We have obtained $\mathcal{T}_\mathrm{sf}$
and $\mathcal{T}_\mathrm{sp}$ from the scattering matrix of the FM barrier,
which does not change below $T_C$.\cite{PhysRevB.62.11812,EshInTanRev} As the 
Josephson current 
is expected to be proportional to the square of the TSC gap magnitude, we
introduce an angular weighting of the transmissivities to account for the 
direction dependence of the gap.  We therefore define the integrated
transmissivity   
\begin{align}\label{quantityweight}
T_\text{int} = \frac{\int_{-\pi/2}^{\pi/2}d\theta\,
  {\cal{T}}_\mathrm{sf}\widetilde{\Delta}_\theta^2}
  {\int_{-\pi/2}^{\pi/2}d\theta\,({\cal{T}}_\mathrm{sp}+{\cal{T}}_\mathrm{sf})
  \, \widetilde{\Delta}_\theta^2}
\end{align}
with
\begin{align}
\widetilde{\Delta}_\theta=
\left\{\begin{array}{cl}
\sin\theta & \quad\text{for } p_y \text{-symmetry}, \\
\cos\theta & \quad\text{for } p_z \text{-symmetry}, \\
1 & \quad\text{for } p_z+ip_y \text{-symmetry}. \end{array}\right.
\end{align} 
Spin-flip
tunneling is dominant for $T_\text{int}>0.5$, whereas for
$T_\text{int}<0.5$
spin-preserving tunneling is most important. As seen in Fig.\
\ref{critcurrentweight}, the barrier widths for which $T_\text{int}=0.5$
are in very good agreement with the sharp minima of 
the critical current at $\alpha=0$ and thus with the 0-$\pi$ transitions. This
shows that Eq.\ (\ref{pertubation_current}) gives a good description of the
relevant physics if one includes the $L$- and $\lambda$-dependence of the
transmissivities.

In Fig.~\ref{critcurrentweight} we also see that the convergence of
$T_{\text{int}}$ towards $0.5$ with increasing $L$ is much faster for
$p_y$-symmetry compared to the other cases. Due to the $p_y$-wave form
factor, shallow incident angles dominate the integral
in~Eq.~(\ref{quantityweight}), which implies averaging over quasiparticle
trajectories corresponding to a large range of effective barrier
widths. Since the probability of spin-flip transmission oscillates with the
effective barrier width, increasing $L$ rapidly leads to equal integrated
probabilities of spin-flip and spin-preserving transmission.
In contrast, $T_{\text{int}}$ decreases more slowly for $p_z$-symmetry
since the weighting favors normal incidence, and thus the average is over a
smaller range of effective barrier widths. The angle-independent weighting for
the $(p_z+ip_y)$-wave case lies between these two
extremes. From Eq.~(\ref{pertubation_current}) we hence deduce that the $\pi$
state in the $p_y$-wave junction is much less robust to increasing
$L$ than for the other cases, consistent with Fig.\ \ref{opimaps1}. 
Since increasing $\lambda$ decreases the length scale of the oscillation
between spin-flip and spin-preserving transmission, this argument also
explains the strong suppression of the $\pi$ states in the $p_y$-wave junction
at moderate to strong polarizations in the ($\alpha$, $\lambda$) maps
(Fig.\ \ref{opimaps2}).

The loss of every second peak in the $\alpha=0$ critical current upon
increasing $\alpha$ to
$\pi/4$ can be 
understood from Eq.~(\ref{pertubation_current}) as resulting from a vanishing
spin-flip contribution. As evidenced by the plots of $T_\text{int}$
in~Fig.\ (\ref{critcurrentweight}) the relative strength
of $\cal{T}_\mathrm{sf}$ and $\cal{T}_\mathrm{sp}$ oscillates with $L$ and
$\lambda$. By removing the spin-flip contribution at 
$\alpha=\pi/4$, the variation of the spin-preserving transmissivity on these
quantities is evidenced in the critical current. Further support for
this interpretation comes from the observation that every second maximum in
the $\alpha=0$ critical current also corresponds to a maximum in
$T_{\text{int}}$, i.e., strong spin-flip tunneling and weak spin-preserving
tunneling. 

\begin{figure}[t]
   \begin{center}
    \includegraphics[width=\columnwidth]{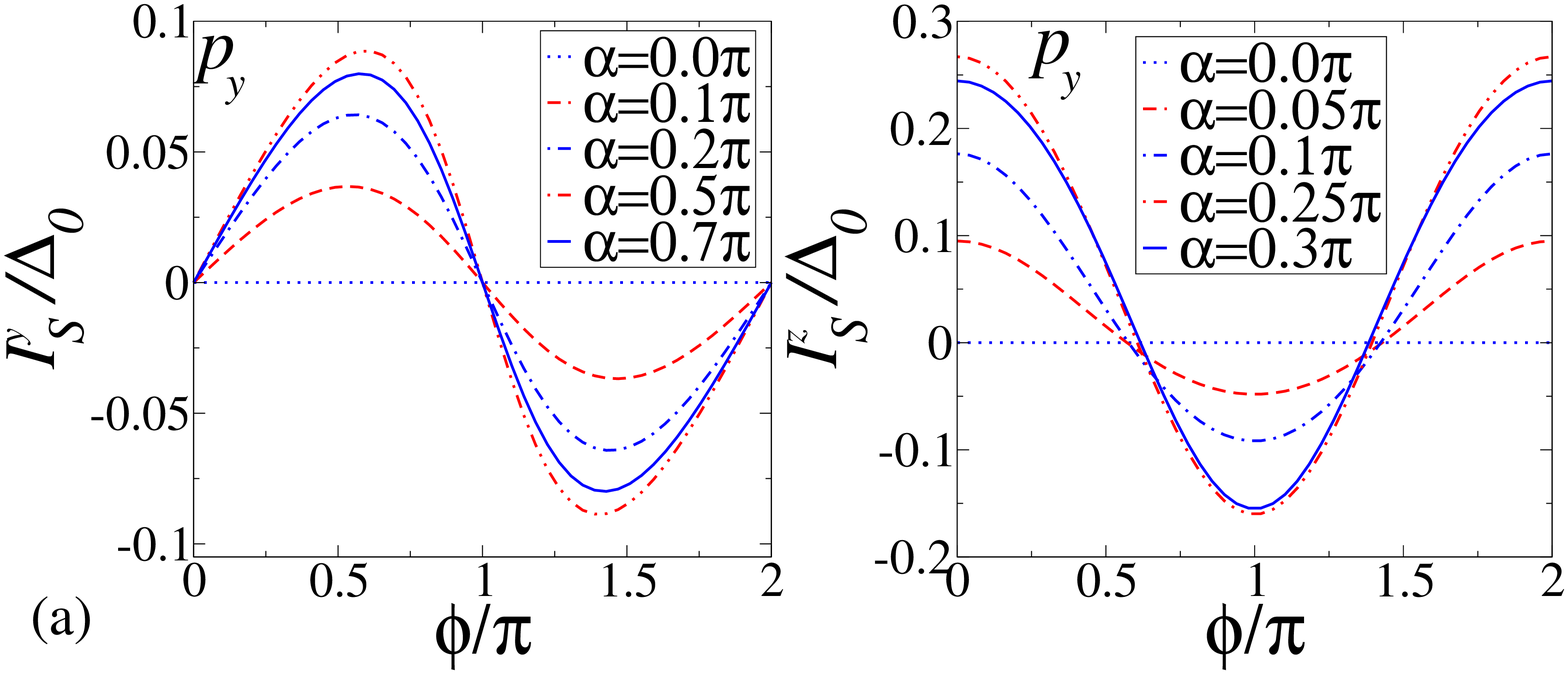}
    \end{center}
   \caption{Spin-current components (a) $I_S^y$, which is due to spin filtering,
   and (b) $I_S^z$, due to spin flip, as functions of the phase difference $\phi$.
   In both panels we set $L=10\,k_F^{-1}$, $\lambda=0.3$, and $T=0.4\,T_C$.
   }
   \label{spindemo}
 \end{figure}

\subsection{Spin current}

\begin{figure*}[t]
   \begin{center}
   \includegraphics[clip,width=2\columnwidth]{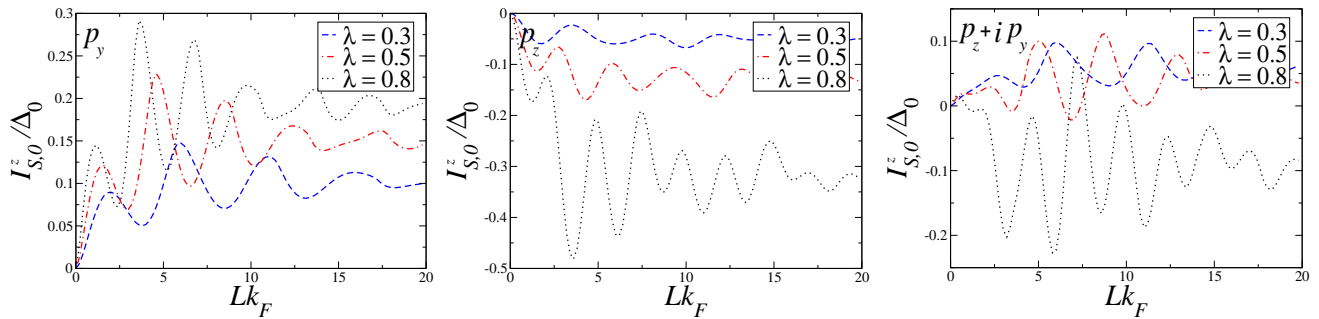}
   \end{center}
   \caption{$\phi$-independent contributions to be spin current as functions of
   the barrier width $L$ for all considered gap symmetries and various exchange
   splittings $\lambda$. In all panels we set
   $\alpha=0.25\,\pi$ and $T=0.4\,T_C$.}\label{new_reflectivities}
\end{figure*}

\begin{figure*}[t]
   \begin{center}
   \includegraphics[clip,width=2\columnwidth]{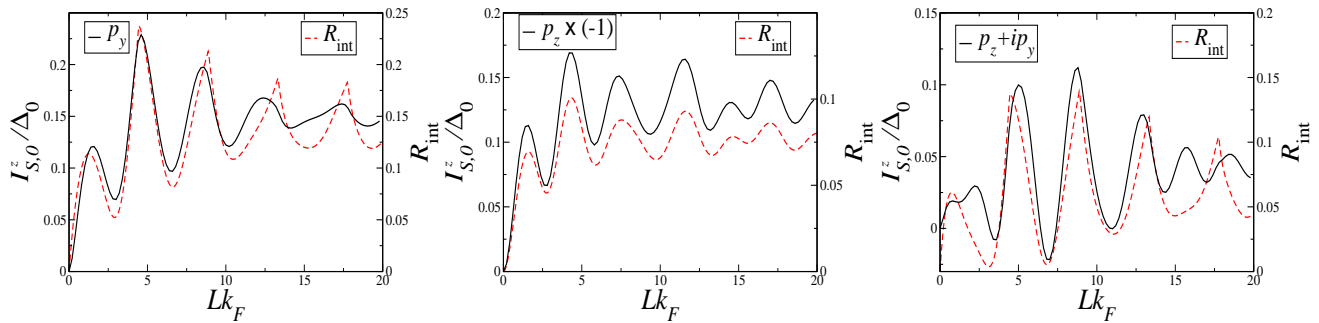}
   \end{center}
   \caption{$\phi$-independent part of the $z$ component of the spin
   current (black solid line, left axis) and integrate spin-flip reflectivity
   $R_\text{int}$  (red dashed line, right axis) as functions of the
   barrier width $L$ for all considered gap symmetries. For $p_z$-wave 
   the current is multiplied
   by $-1$ for better comparability. We set $T=0.4\,T_C$ and $\lambda=0.5$ in all
   panels.}\label{weighted_reflectivities}
\end{figure*}

We find that the polarization of the spin current always lies in the
$y$-$z$ 
plane. The $y$ and $z$ components are plotted in Fig.\ \ref{spindemo}
for the $p_y$-wave case; the results are qualitatively identical for the
other orbital symmetries. We show
below that these two polarizations are due to different mechanisms, as proposed
by Brydon \textit{et al.}\cite{JPSP}

The $y$ component of the spin current shows the $\alpha$ and $\phi$ dependence
expected for a spin-filtering mechanism.\cite{PhysRevB.77.104504,JPSP} It occurs
when the magnetic moment of the barrier has a component perpendicular to both
$\mathbf{d}$-vectors of the TSCs. The spin-$\up$ and spin-$\down$
states with respect to this direction are not mixed by scattering at the barrier
but do experience different barrier
transmissivities, leading to spin filtering of the
current. For a $\delta$-function barrier, the filtering effect requires a
non-magnetic scattering potential at the interface. This is not the case for a finite
magnetic barrier due to the spin splitting of the Fermi surface. Specifically,
minority-spin quasiparticle propagation is suppressed in the FM region above
the critical injection angle $\theta_{\text{crit}}$ defined in Eq.\ (\ref{critangle}),
which is equivalent to a lowered transmissivity of the minority spins compared
to the majority spins, giving rise to spin filtering.

The $z$ component of the spin current shows the $\alpha$ and
$\phi$ dependence expected for a spin-flipping
mechanism.\cite{PhysRevLett.103.147001,JPSP} This occurs due to the
spin-dependent phase shift acquired by a triplet Cooper pair when it
undergoes a 
spin-flip during the tunneling across the barrier.
This mechanism requires a component of the magnetic moment parallel to the
$\mathbf{d}$-vectors of both TSCs. In the perturbation theory for a
$\delta$-barrier junction the $z$ 
component of the spin current satisfies \cite{PhysRevLett.103.147001}
\begin{align}\label{pertubation_spincurrent}
I_S^{z,\text{pert}} \propto ( T_\mathrm{sf}\cos\phi + \gamma R_\mathrm{sf} )\,
  \sin 2\alpha \, ,
\end{align}
where $T_\mathrm{sf}$ and $R_\mathrm{sf}$ are tunneling matrix elements of
spin-flip tunneling and spin-flip reflection and $\gamma$ is an
orbital-dependent 
factor connected to the phase shift experienced by specularly reflected
quasiparticles. For the $p_y$ junction, there is no phase shift upon reflection
and it is found that $\gamma=1$. In a $p_z$ junction, reflected Cooper pairs
experience a $\pi$ phase shift leading to a factor of $\gamma=-1$ and in the
more complicated case of a $p_z+ip_y$ junction the phase shift of the Cooper
pairs depends on their incident angle as $\pi-2\arctan(k_y/k_z)$ so that
$-1<\gamma<0$.

We obtain the $\phi$-independent contribution $I_{S,0}^z$ by performing
a Fourier transformation 
of $I_S^z(\phi)$. Figure \ref{new_reflectivities} shows
that $I_{S,0}^z$ oscillates with increasing
barrier width $L$ for all considered gap symmetries. These oscillations are
due to interference between quasiparticles reflected off the two interfaces. 
The sign difference between $I_{S,0}^z$ for the
$p_y$-wave and $p_z$-wave 
cases agrees with previous
results;\cite{PhysRevLett.103.147001,PhysRevB.80.224520,PhysRevB.83.180504} in
particular, we find that the
sign of the $\gamma$ factor in 
Eq.\ (\ref{pertubation_spincurrent}) for the finite width barrier is the same
as for the $\delta$-function barrier.
The $(p_z+ip_y)$-wave case shown in 
Fig.\ \ref{new_reflectivities} is intermediate between these other cases, and
we find that $\gamma$ can 
be positive or negative depending on 
$\lambda$ and $L$, in contrast to the $\delta$-function results for which the
sign was found to be always negative.\cite{PhysRevLett.103.147001}
Specifically, in Fig.\ \ref{new_reflectivities} we see that for 
  $(p_z+ip_y)$-wave symmetry $I_{S,0}^z$ is always positive at $\lambda=0.3$,
at $\lambda=0.5$ it is negative for some values of $L$, and it is
mostly negative at $\lambda=0.8$.

In analogy to the procedure for the charge current, we demonstrate the
appropriateness of Eq.\ (\ref{pertubation_spincurrent}) to describe the
relevant physics, by showing that the oscillations of $I^{z}_{S,0}$ are due to
oscillations of the spin-flip reflectivity 
$\mathcal{R}_\mathrm{sf}$. 
The parameter $R_\mathrm{sf}$ in Eq.\ (\ref{pertubation_spincurrent})
is expected to be roughly proportional to the reflectivity
$\mathcal{R}_\mathrm{sf}$.\cite{PhysRevLett.103.147001}
Since increasing the magnetization of the barrier makes it more likely for
particles to be spin-flip reflected, i.e., $\mathcal{R}_\mathrm{sf}$ should be
enhanced, we see that $I^{z}_{S,0}$ increases with
$\lambda$ 
for all gap symmetries,
see Fig.\ \ref{new_reflectivities}. We have also compared $I^{z}_{S,0}$ to the
weighted and $\theta$-integrated spin-flip reflectivity 
\begin{align}
R_\text{int} = \int_{-\frac{\pi}{2}}^{\frac{\pi}{2}}d\theta\,
  \mathcal{R}_\mathrm{sf}(\theta)\widetilde{\Delta}^2_{\theta}
\end{align}
for a finite-barrier normal metal-FM-normal metal junction. The weighting is
carried out in the same way as for the charge current except for the
$p_z+ip_y$-wave 
case, where it is necessary to account for the $\theta$-dependent phase shift
of quasiparticles reflected at the interface. Thus we weight the 
normal metal-FM-normal metal reflectivity by
$\widetilde{\Delta}_\theta^2=\sin^2 \theta-\cos^2\theta$.
Fig.\ \ref{weighted_reflectivities} shows that the variation of $I_{S,0}^z$ is in
very good agreement with $R_\text{int}$,  
especially for smaller values of $L$ and $\lambda$. We also find that
the sign change in $I_{S,0}^z$ for the $(p_z+ip_y)$-wave case reflects an
important change in the nature of the spin-flip
reflection:~\cite{PhysRevB.80.224520,PhysRevB.83.180504} the positive
$I_{S,0}^z$ at small $\lambda$ indicates
$p_y$-like behaviour, where tunneling of quasiparticles with high incident
angles $\theta\approx\pi/2$ is dominant. At higher values of $\lambda$,
however, the transmission of 
quasiparticles with lower incident angles $\theta\approx 0$ is favored, leading
to a $p_z$-like behavior and hence negative $I_{S,0}^z$. We hence conclude
that when the $L$ and $\lambda$ dependence of the transmissivity and
reflectivity are taken into account, Eq.\ (\ref{pertubation_spincurrent})
qualitatively captures the observed behavior of the spin current.

\section{Summary and Discussion}

In this paper we have studied the charge and spin currents in a
TSC-FM-TSC junction. 
For the charge current, we have found that transitions from a $0$- to a
$\pi$-state are induced by rotating the magnetization of the FM barrier in the
plane 
parallel to the \textbf{d}-vectors of both TSCs or by varying its
magnitude. These effects have already been predicted for a
$\delta$-function barrier.\cite{PhysRevLett.103.147001}
In addition, we have found that changing the barrier
width can also cause a $0$-$\pi$ transition. In all cases, $0$-$\pi$ transitions
are accompanied by sharp minima of the critical current, which can thus serve
as diagnostics. For the case of $(p_z+ip_y)$-wave symmetry, our results for the
critical charge current agree with Rahnavard \emph{et al.}\cite{Rahnavard}\
However, we have gone beyond their work by considering two additional
gap symmetries, by obtaining $0$-$\pi$ phase maps, and by extending the
range of the exchange splitting. Furthermore, we have explained the
  origin of the charge current in terms of the variation of the 
spin-flip and spin-preserving transmissivities:
for all considered gap symmetries and barrier parameters, the latter dominates
over the former in a $0$-state, whereas the opposite is true for a
$\pi$-state. This result is consistent with lowest-order perturbation theory.

The spin currents in the finite-barrier case can be classified in the same
manner as for the $\delta$-function barrier, i.e., we find spin currents due to
spin filtering and to spin flipping. Unlike for the $\delta$-function
barrier, however, the spin-filtering effect can occur in the absence
of a finite charge scattering potential because of the spin splitting of
the Fermi surface in the FM barrier. 
For the spin current in the $(p_z+ip_y)$-wave case we find very different
polarizations and 
phase dependences compared to Rahnavard \emph{et
  al.}\cite{Rahnavard}\ Specifically, the authors of
Ref. \onlinecite{Rahnavard} found a spin current
polarization parallel to the 
$\mathbf{d}$-vector, which is 
inconsistent with the fact that the spin of the Cooper
pairs is always perpendicular to the $\mathbf{d}$-vector. The origin of this
discrepancy might be that,
in contrast to our work, in Ref. \onlinecite{Rahnavard} the spin current was
evaluated in the FM region, where the spin current is not a conserved quantity
and is position dependent.

Overall, our results show that approximating the barrier as a
$\delta$-function 
gives a qualitatively correct account of the Josephson currents in a TSC-FM-TSC
junction. This demonstrates the robustness and, since a metallic ferromagnetic
barrier is realizable in a Josephson junction, the experimental relevance of
the previously predicted
phenomena.\cite{PhysRevLett.103.147001,Asano,JPSP} In particular, the
sign reversal of the current upon varying the orientation of the
magnetic moment is found to be a robust test of triplet
superconductivity. The 
inclusion of a finite barrier width is nevertheless crucial for a
quantitatively realistic description, as the barrier width can by
itself control 0-$\pi$ transitions of the junction.

\acknowledgments

We have profited from discussions with G. Annunziata and T. Yokoyama.


\begin{thebibliography}{99}
\bibitem{meaculpa}Note that we have corrected an error in the overall sign of 
  $I^{z}_{S}(\phi)$ that was present in Refs. ~\onlinecite{PhysRevLett.103.147001,PhysRevB.80.224520}.
\end{thebibliography}
\end{document}